# Thickness-dependent bulk properties and weak anti-localization effect in topological insulator $Bi_2Se_3$

Yong Seung Kim[1,2], Matthew Brahlek[1], Namrata Bansal[3], Eliav Edrey[1], Gary A. Kapilevich[1], Keiko Iida[4], Makoto Tanimura[4], Yoichi Horibe[1], Sang-Wook Cheong[1] and Seongshik Oh[1,*]

[1] Department of Physics & Astronomy, The State University of New Jersey, Piscataway, New Jersey 08854, U.S.A.

[2] Department of Physics and Graphene Research Institute, Sejong University, Seoul 143-747 (Korea)

[3] Department of Electrical and Computer Engineering, Rutgers, the State University of New Jersey, Piscataway, New Jersey 08854 (USA)

[4] Research Department, Nissan Arc, Ltd., Yokosuka, Kanagawa 237-0061 (Japan)

[*] ohsean@physics.rutgers.edu

(Dated: May 11, 2011)

## Abstract

**We show that a number of transport properties in topological insulator (TI) $Bi_2Se_3$ exhibit striking thickness-dependences over a range of up to five orders of thickness (3 nm – 170 µm). Volume carrier density decreased with thickness, presumably due to diffusion-limited formation of selenium vacancies. Mobility increased linearly with thickness in the thin film regime and saturated in the thick limit. The weak anti-** **localization effect was dominated by a single two-dimensional channel over two decades of thickness. The sublinear thickness-dependence of the phase coherence length suggests the presence of strong coupling between the surface and bulk states.**

Over the past few years, topological insulators (TI) have emerged as a new platform for coherent spin-polarized electronics. TIs are predicted to have an insulating bulk state and spin-momentum-locked metallic surface states [1-9]. This spin-momentum-locking mechanism and their band structure topology are predicted to prevent the surface metallic states from being localized due to backscattering. Among the TIs discovered so far, $Bi_2Se_3$ is considered the most promising due to its near-ideal surface-band structure [7], and its predicted surface states have been confirmed by a number of surface-sensitive probes [5-6, 10-12]. However, transport properties in this material have been far from the theoretical prediction of metallic surface states with insulating bulk states; instead, the bulk state has always turned out to be metallic and dominated the total conductance [12-14]. Considering that the bulk and surface transport properties should exhibit very different thickness-dependences, thickness-dependent transport studies can provide insights on both the bulk and surface states. Here we undertook extensive thickness-dependent studies of the transport properties in $Bi_2Se_3$ covering up to five orders of sample thickness, from which we made a number of critical findings that have been hidden in previous studies of $Bi_2Se_3$.

For this study, we grew $Bi_2Se_3$ films with atomically-sharp interfaces on undoped Si(111) substrates (10 x 10 x 0.5 $mm^3$) by molecular beam epitaxy (MBE), covering more than three orders of thickness from 3 quintuple layers (QL, 1 QL ≈ 1 nm) through 3,600

QL. From a technological point of view, silicon is the most important substrate for electronic applications. However, interfaces between $Bi_2Se_3$ thin films and Si substrates have been plagued by disordered interfacial layers [14-15]. In order to solve this problem, we developed a two-step growth scheme and obtained high-quality second-phase-free $Bi_2Se_3$ films on Si substrates with atomically sharp interfaces [16]. Such an atomically-sharp interface is crucial for reliable thickness-dependent studies for very thin samples. A cross-section TEM image of a $Bi_2Se_3$ film in Fig. 1(a) shows an atomically sharp interface between the film and the Si substrate. The sharp reflection high energy electron diffraction (RHEED) pattern in the inset represents the high crystallinity of the $Bi_2Se_3$ film. Figure 1(b) shows X-ray diffraction (XRD) patterns for three different thicknesses. The peaks observed in the $\theta$-$2\theta$ scan are consistent with the c-axis oriented $Bi_2Se_3$ phase. During the growth of the film, the gap between each line in the RHEED pattern, which is inversely proportional to the in-plane lattice constant of the film, changed to that of the bulk lattice constant during the first QL growth. This immediate lattice relaxation is attributed to the weak van der Waals type bonding between the film and the substrate.

The transport measurements were carried out with the standard van der Pauw method with indium contacts on four square corners in a cryostat with magnetic field up to 9 T and a base temperature of 1.5 K; measurement errors due to the contact geometry are estimated to be less than 10%. For all films except for 3 QL, the R vs. T curves in Fig. 2(a) showed metallic behaviour at high temperature, but as the temperature dropped below 30 K the resistance tended to increase slightly either due to strong electron-electron interaction in the 2D limit [17] or due to an impurity band in the bulk [18]. The following measurements were all taken at a fixed temperature of 1.5 K with each sample exposed to air for less than 20 minutes to minimize air contamination; at this temperature, the undoped Si substrates were completely insulating and did not contribute to the transport measurements.

It is well known that the mobility of conventional thin films depends on the film thickness as $\mu(t) = \mu_\infty/(1+2(\lambda/t)(1-p))$, where $\mu_\infty$ is the mobility of the film when the thickness, $t$, is much larger than the mean free path, $\lambda$, with $p$ representing the fraction of carriers reflecting specularly from the surface [19]; this mobility drop originates from the reduction in the effective mean free path caused by diffuse scattering from the surfaces. Figure 2(b) shows that the mobility vs. thickness data, obtained from Hall effect measurements, are in good agreement with this standard theory with $\mu_\infty \approx 3,000$ cm$^2$/Vsec and $\lambda(1-p) \approx 70$ nm, except for the 3 QL data marked with a triangle. If the bulk of the film is insulating and the transport is entirely confined to the surfaces, the mobilities should be thickness-independent except for very thin samples where quantum tunnelling between the top and bottom surfaces can degrade the surface states [20-22]. The very observation of the conventional thickness-dependence implies that the observed mobilities are dominated by the bulk instead of the surface transport.

While the thickness-dependence of the mobilities can be well understood by the standard surface scattering theory, the thickness dependence of the volume and sheet carrier densities plotted in Figs. 2(c)-(d) is puzzling and unexpected. The volume carrier density (defined as the total sheet carrier density, obtained from the Hall measurement, divided by the sample thickness) decreases monotonically as the thickness increases and scales as $t^{-0.5}$ over three orders of thickness range, from $5.3 \times 10^{19}$ cm$^{-3}$ for 3 QL to $1.6 \times 10^{18}$ cm$^{-3}$ for 3,600 QL. Little change occurred in these values even after the films were annealed in high selenium vapour pressures, up to six orders of magnitude higher than the normal growth conditions; this implies that the observed carrier densities are close to the absolute minimum values that are experimentally achievable in these pure $Bi_2Se_3$ films. We examined published data obtained from single crystal samples [18], and surprisingly these data points fell on the same curve, extending the trend up to five orders of thickness (3 nm - 170 µm) with $2\times10^{17}$ cm$^{-3}$ for 170 µm thick single crystal [18]. Considering that two completely different sample formation processes, one through MBE growth and the other through peeling of bulk samples [18], result in the same thickness-dependence suggests some fundamental mechanism behind this anomalous thickness-dependence of the carrier densities. Another view of this anomaly is through the thickness-dependence of the sheet carrier density. If we assume constant bulk volume carrier density ($n_{bv}$) and constant surface carrier density ($n_s$), the total sheet carrier density ($n_{sheet}$) should scale linearly with the sample thickness (t) such that $n_{sheet} = n_s + n_{bv}t$. In Fig. 2(d), the theoretical curve with $n_s = 2\times10^{13}$ cm$^{-2}$ and $n_{bv} = 1\times10^{18}$ cm$^{-3}$ was plotted for comparison with the experimental data. The observed data can in no way be explained by this simple model, and instead the total sheet carrier density scales as $t^{0.5}$. This implies that the bulk volume carrier density is not constant but varies monotonically with its thickness over five orders of magnitude. Considering that either the TI surface state [20] or the surface band-bending effect [23] can never extend more than tens of nanometers while the observed anomaly extends far beyond the micrometer scale, associating this observation with such an electronic mechanism seems unphysical. Because the volume carrier density mainly originates from the selenium vacancies [24], this observation nominally implies that the volume density of selenium vacancies gradually increases as samples get thinner. The formation of selenium vacancies through diffusion seems to continuously occur even at room temperature, as confirmed by other group with X-ray photoelectron spectroscopy [25]. These observations suggest that formation of selenium vacancies in $Bi_2Se_3$ is thermodynamically favourable at room temperature, yet it occurs through a slow diffusion process, which is inevitably thickness-dependent. This also implies that the measured carrier density of a sample depends on the time between sample fabrication and measurement, and so in order to maintain consistency, almost all samples reported here were measured on the day they were fabricated. In other words, the exact thickness dependence of the carrier densities, especially in the thin limit, may depend on when they are measured.

The magneto-resistance (MR) measurements provide another means to probe the TI properties. In Figs. 3(a)-(b), the MR, defined as (R(B)-R(0))/R(0), in the high field regime is dominated by the parabola-like ($B^2$) dependence originating from the Lorentzian deflection of carriers under perpendicular magnetic field [26] (recall that the electron executes cyclotron orbits, thereby shortening the mean-free-path, and thus increasing the resistance). The fact that this $B^2$-dependence is more pronounced in thicker samples suggests that it is a bulk-dominated effect.

In the low field regime (<0.5 T) for thinner samples, the magneto-conductance (MC), as shown in Fig. 4(a), decreases sharply as the magnetic field is increased, which is typical of the weak anti-localization (WAL) effect [27-28]. This WAL effect is the result of the strong spin-orbit coupling, which puts backscattering at the minimum when there is no magnetic field, due to the time-reversal symmetry. As magnetic field increases, thus breaking the time reversal symmetry, backscattering increases and leads to a sharp reduction in conductance. The low field MC, $\Delta G(B) = G(B)-G(0)$, can be well fitted to the standard Hikami-Larkin-Nagaoka (HLN) theory for WAL [29]: $\Delta G(B) = A(e^2/h)[ln(B_\phi/B) - \Psi(1/2+B_\phi/B)]$, where $A$ is a coefficient predicted to be $1/(2\pi)$ for each 2D channel, $B_\phi$ is the de-phasing magnetic field, and $\Psi(x)$ is the digamma function. The de-phasing magnetic field is related to the phase coherence length $l_\phi$ via [17, 30] $B_\phi = \hbar/(4el_\phi^2)$. Figure 4b-c shows the fitting parameters as a function of thickness for 3-100 QL. Except for 3 QL, the fitting parameter $A$ remains approximately constant around $1/(2\pi)$ while the parameter $l_\phi$ monotonically increases as samples get thicker. The parameter $A$ being closer to $1/(2\pi)$ than $1/\pi$ nominally implies that the WAL effect over 5 - 100 QL is dominated by a single 2D channel. There are two possibilities for this observation. The first is that the WAL effect originates entirely from the reduced dimensionality of the metallic bulk state without any contribution from the surface states. The other is that the surface states contribute but they couple strongly with the metallic bulk state and behave together as a single 2D system. We will show below that the thickness-dependence analysis of $l_\phi$ supports the latter. The WAL signal became too small to be of quantitative relevance as the thickness increased beyond 100 QL, but the spike in $\Delta G$ was still visible up to 3,600 QL as shown in the inset of Fig. 3(b), suggesting the robustness of the WAL effect. Below 5 QL, the WAL behaviour significantly degrades as the oppositely spin-polarized top and bottom surfaces start to couple strongly through quantum tunnelling [20-22].

Figure 4(c) shows that the phase coherence length $l_\phi$ scales as $t^{0.7}$. If the WAL effect originated entirely from the geometrically-confined bulk state of the film without any surface state contribution, then $l_\phi$ should scale linearly with thickness in the thin film limit, just as the bulk-dominated mobility – a quantity proportional to a mean free path - presented in Fig. 2(b) scales almost linearly with the thickness over 5 – 100 QL. Therefore, the sublinear thickness-dependence of $l_\phi$ suggests that there should be a surface state contribution to the WAL signal and that the observed single channel effect must be due to strong coupling between the surface and bulk states. Now if the bulk of the film were insulating, then $l_\phi$, being a surface property, had to be independent of the thickness—the surfaces should not change when the thickness changes. However, with conducting bulk states, the surfaces and bulk can interact with one another and lead to a thickness-dependent scattering mechanism. In other words, the thickness-dependent coherence length is a natural result of the metallic bulk states. According to this analysis, the thickness dependence of $l_\phi$, denoted by $\alpha$ in $l_\phi \sim t^\alpha$, could be used as a figure-of-merit to tell how close certain TI materials are to an ideal TI; $\alpha$ should be zero for an ideal TI with the insulating bulk state, one for a topologically-trivial strongly spin-orbit-coupled metal, and between zero and one for a non-ideal TI with the metallic bulk state, approaching zero as the material gets closer to an ideal TI.

In summary, extensive thickness-dependent studies of the $Bi_2Se_3$ transport properties have led to a number of unexpected findings. The volume carrier density, which is commonly assumed to be thickness-independent, decreased by more than two orders of magnitude over five orders of thickness, suggesting that selenium diffusion is highly thickness-dependent and active even at room temperature. The mobility increased linearly with thickness in the thin film regime and saturated as the samples got thicker, suggesting that the surface scattering effect limits the mean free path in the thin limit. The WAL effect was dominated by a single 2D channel over two decades of thickness. The sublinear thickness-dependence of the phase coherence length supports the presence of the surface states and provides a figure-of-merit characterizing the level of interaction between the surface and bulk states. Our observations suggest that interactions between the bulk and surface states have profound effects on their transport properties.

We thank Keun Hyuk Ahn for reviewing the manuscript before submission. This work is supported by IAMDN of Rutgers University, National Science Foundation (NSF DMR-0845464) and Office of Naval Research (ONR N000140910749).

**Figure Legends**

**FIG. 1 (color online). Epitaxial growth of $Bi_2Se_3$ film.**

(a) Cross-section TEM image of 32 QL film shows atomically sharp interface between $Bi_2Se_3$ and a Si substrate. Inset: RHEED pattern of the $Bi_2Se_3$ film. (b) High-resolution X-ray diffraction pattern of three different films.

**FIG. 2 (color online). Thickness dependence of the transport properties of $Bi_2Se_3$.**

(a) Resistance of $Bi_2Se_3$ films as a function of temperature with thickness ranging from 3 to 3,600 QL. Data above ~150 K are not shown here because they are affected by the parallel conduction of thermally excited carriers in the un-doped silicon substrates. (b) Mobility, (c), (d) Carrier density of electrons in $Bi_2Se_3$ films obtained by Hall measurement at 1.5 K. The solid curve in (b) is $\mu(t) = 3,000/(1+140/t)$, and the straight lines in (c) and (d) are illustrative guides.

**FIG. 3 (color online). Normalized Magnetoresistance.**

The magnetic field dependence of resistance at 1.5 K of (a), thin film from 3 to 32QL and (b) thick film from 60 to 3,600 QL. Deep cusp in low field regime is characteristic of the WAL effect. Parabolic field dependence is dominant in thick films. Inset of (b): zoomed-in view of the 3,600 QL data near zero field showing robustness of the WAL effect.

**FIG. 4 (color online). Weak anti-localization effect.**

(a) The HLN fitting of the change in conductance in low field regime from 3 to 100 QL. (b) Above 5 QL, the coefficient *A* is almost thickness-independent, corresponding to a single channel WAL contribution. (c) The phase coherence length increases as $t^{0.7}$ with the sample thickness.

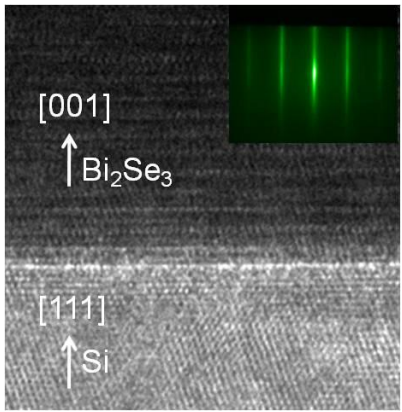

Fig. 1(a)

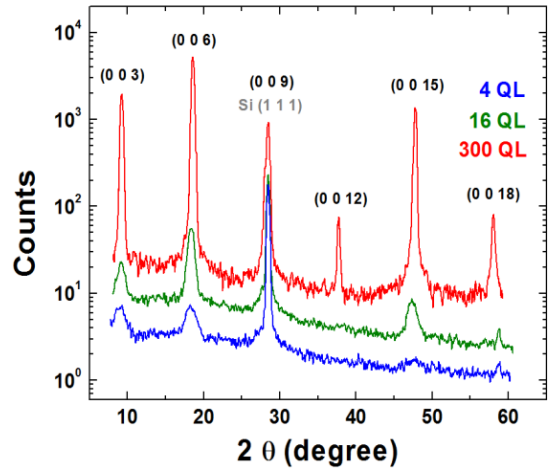

Fig. 1(b)

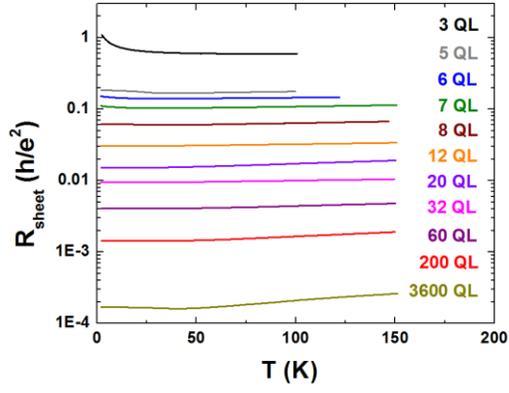

Fig. 2(a)

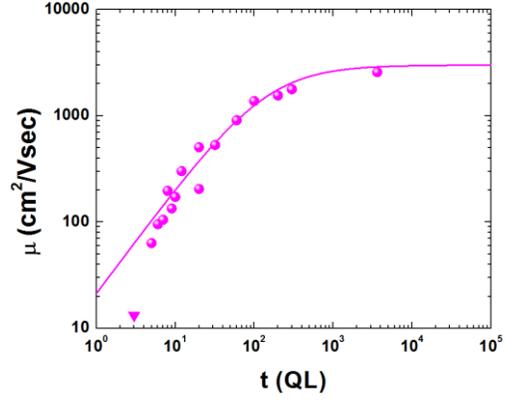

Fig. 2(b)

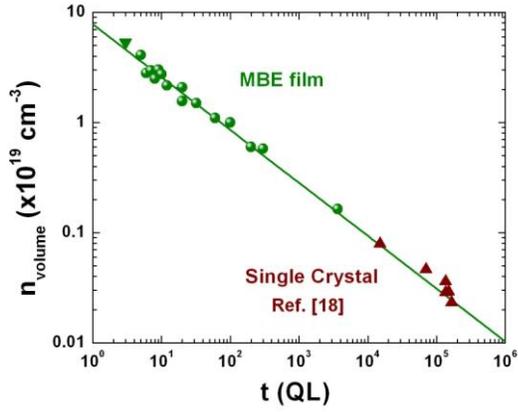

Fig. 2(c)

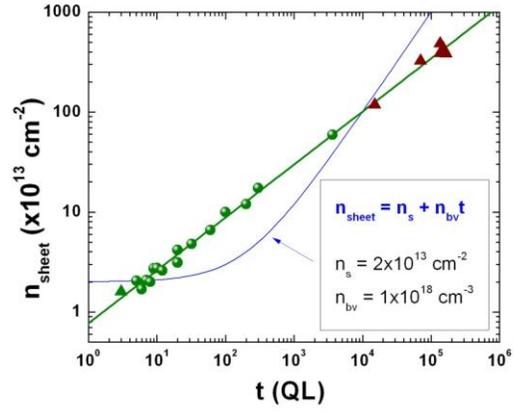

Fig. 2(d)

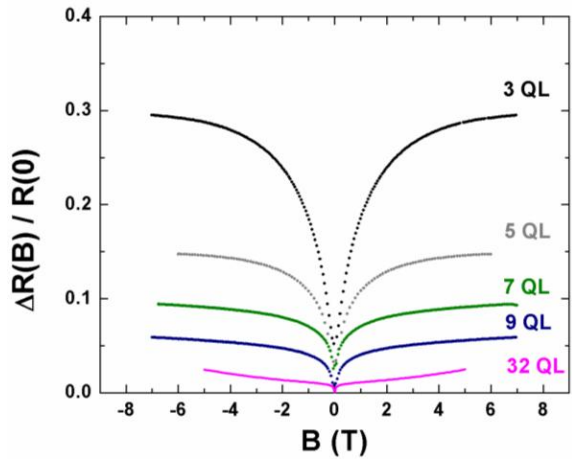 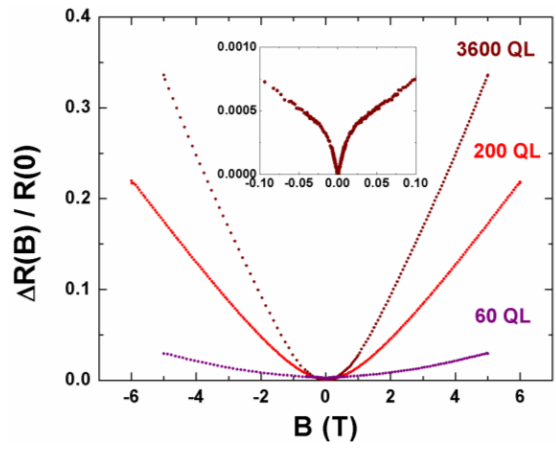

Fig. 3(a)  Fig. 3(b)

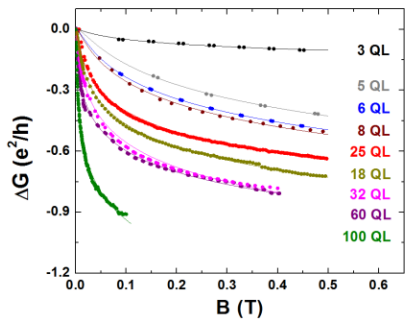 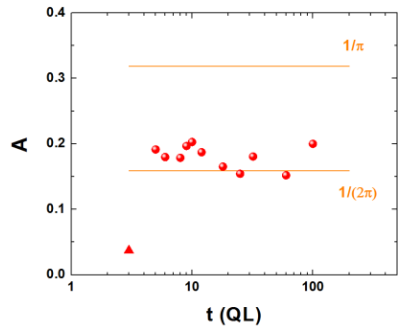 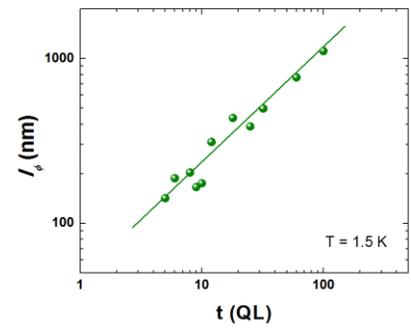

**Fig. 4(a)**        **Fig. 4(b)**        **Fig. 4(c)**